\documentclass[]{spie} 
 
\usepackage[]{graphicx}
\usepackage{amsmath}
\usepackage{revsymb}
\usepackage{amssymb}
\usepackage{bm}
\usepackage{overpic}
\usepackage[svgnames]{xcolor}

\title{Ultra-low emittance beam generation using two-color ionization
injection in a {\bf \LARGE CO$_2$} laser-driven plasma accelerator}

\author{C. B. Schroeder\supit{a}, C. Benedetti\supit{a}, S. S.
Bulanov\supit{a}, M. Chen\supit{b}, E. Esarey\supit{a},
C.~G.~R.~Geddes\supit{a}, J.-L. Vay\supit{a}, L.-L. Yu\supit{b}, W. P.
Leemans\supit{a} \skiplinehalf \supit{a}Lawrence Berkeley National
Laboratory, Berkeley, California 94720, USA; \\
\supit{b}Shanghai Jiao Tong University, Shanghai 200240, China }

\begin{document} 
  \maketitle 

\begin{abstract}
Ultra-low emittance (tens of nm) beams can be generated in a plasma
accelerator using ionization injection of electrons into a wakefield.
An all-optical method of beam generation uses two laser pulses of
different colors. A long-wavelength drive laser pulse (with a large
ponderomotive force and small peak electric field) is used to excite a
large wakefield without fully ionizing a gas, and a short-wavelength
injection laser pulse (with a small ponderomotive force and large peak
electric field), co-propagating and delayed with respect to the pump
laser, to ionize a fraction of the remaining bound electrons at a
trapped wake phase, generating an electron beam that is accelerated in
the wake. The trapping condition, the ionized electron distribution,
and the trapped bunch dynamics are discussed. Expressions for the beam
transverse emittance, parallel and orthogonal to the ionization laser
polarization, are presented. An example is shown using a 10-$\mu$m
CO$_2$ laser to drive the wake and a frequency-doubled Ti:Al$_2$O$_3$
laser for ionization injection.
\end{abstract}



\section{INTRODUCTION}
\label{sec:intro}  

Plasma-based accelerators \cite{Esarey09} can produce extremely large
accelerating gradients, enabling compact sources of high-energy beams.
Rapid experimental progress has occurred in the field of laser-driven
plasma accelerators, and electron beams accelerated to multi-GeV
energies have been demonstrated\cite{Wang13,Leemans14} using an
intense laser driving a plasma wave in cm-scale plasmas.
Generation of these electron beams relied on self-injection from the
background plasma in highly-nonlinear plasma waves \cite{Benedetti13}.
In the regime where background plasma electrons are self-trapped,
experiments show that beams with sub-micron normalized transverse
emittance can be produced \cite{Plateau12,Weingartner12}.

\begin{figure}
   \begin{center}
   \begin{tabular}{c}
\includegraphics[]{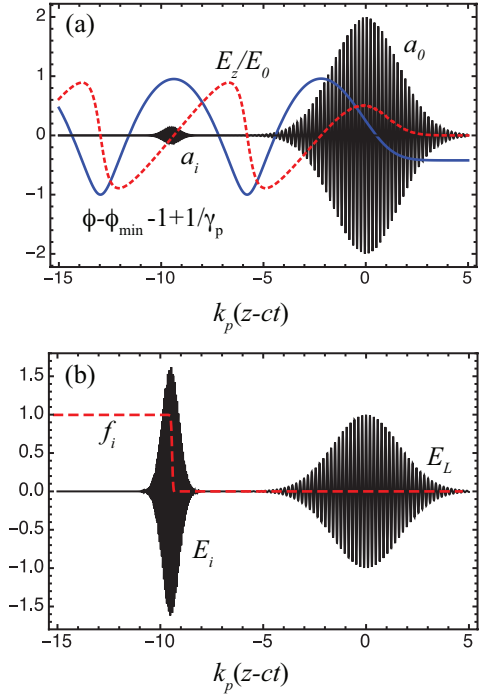}
   \end{tabular}
   \end{center}
   \caption[] 
{(a) Drive laser vector potential $a_0$, ionization laser vector
potential $a_i$, excited plasma wakefield $E_z/E_0$ (red dashed
curve), and trapping condition $0< \phi - \phi_{\rm min} - 1 +
1/\gamma_p$ (blue curve). (b) Drive ($E_L$) and ionization ($E_i$)
laser electric fields (normalized to the peak of $E_L$) and the
fraction of the high-Z gas state ionized $f_i$ (red dashed curve).
\label{fig:1d} }
   \end{figure}

To reduce the electron beam emittance further it has been proposed to
use a laser to ionize electrons at a trapped phase in a plasma wake
that is independently excited by a particle beam \cite{Hidding12} or
an intense laser \cite{Yu14}. The laser-based method \cite{Yu14}
relies on two laser pulses of different colors: a long wavelength
pulse, with large ponderomotive force and small peak electric field,
excites a plasma wake without fully ionizing a high-Z gas; a
short-wavelength injection pulse, with small ponderomotive force and
large peak electric field, co-propagating and delayed with respect to
the wake drive laser, ionizes a fraction of the remaining bound
electrons at a trapping phase of the wake, generating an electron
beam. Figure~\ref{fig:1d} illustrates the basic principle of the
two-color ionization injection method. This two-color, two-pulse
ionization injection concept was first proposed in
Ref.~\citenum{Yu13}. An additional numerical study was performed in
Ref.~\citenum{Xu14b} subsequent to Refs.~\citenum{Yu13,Yu14}.

In this paper we discuss the trapping condition, the distribution of
ionized electrons, and the dynamics of the trapped electron beam. We
present expressions for the transverse emittance, parallel and
orthogonal to the laser polarization, of the trapped electron beam,
valid for beam or laser wakefield drivers.\cite{Schroeder14} For the
all-optical, two-color ionization injection method, it is natural to
consider a 10-$\mu$m CO$_2$ laser as the long-wavelength drive pulse,
generating the plasma wake, followed by a Ti:Al$_2$O$_3$ laser
(frequency-doubled, with 0.4~$\mu$m wavelength) for ionization
injection. Progress in CO$_2$ laser technology has opened the
possibility of sub-ps pulse durations, that would enable efficient
(i.e., resonant, with duration of order the plasma period) plasma
wakefield excitation at plasma densities $\sim 10^{16}$~cm$^{-3}$, and
such laser systems are expected to become available in the next
several years \cite{Pogorelsky14}. We present an example of two-color
ionization injection using a short-pulse, CO$_2$ drive laser pulse and
a frequency-doubled, Ti:Al$_2$O$_3$ injection laser pulse.

\section{TRAPPING PHYSICS} 

\subsection{Trapping condition}

The two-color ionization injection concept relies on a large plasma
wave (or wakefield) excited by a long-wavelength drive laser pulse,
with wavelength $\lambda_0 = 2\pi/k_0$ and normalized vector potential
amplitude $a_0 = eA_0/m_ec^2$, where $e$ and $m_e$ are the electron
charge and mass, respectively, and $c$ is the speed of light. For
efficient wakefield generation the drive pulse should be approximately
resonant with the plasma, with $k_p L_0 \sim 1$, where $L_0$ is the
drive laser length and $\omega_p = k_p c = (4\pi e^2 n_e/m_e)^{1/2}$
is the plasma frequency, with $n_e$ the electron number density.
Co-propagating and delayed with respect to the drive pulse is an
ionization laser pulse, with wavelength $\lambda_i = 2\pi/k_i <
\lambda_0$ and normalized vector potential amplitude $a_i =
eA_i/m_ec^2$. The ionization pulse electric field $E_i = (2\pi m_e
c^2/e)a_i/\lambda_i$ is sufficiently large, $a_i/\lambda_i \gtrsim
a_0/\lambda_0$, to ionize remaining bound electrons in the gas, that
may be trapped in the wakefield.

Trapping requires that the electrons are ionized at the proper wake
phase and that the wake amplitude is sufficiently large. Behind the
drive laser pulse, and assuming $a_i^2(\psi_i)\ll 1$, this condition
may be expressed as \cite{MinChen12}
\begin{equation}
  1-\gamma_p^{-1}  \leq  \phi (\psi_i)- \phi_{\rm min} 
,
\label{eq:trap-cond}
\end{equation}
where $\psi_i = k_p \xi = k_p (z-ct)$ is the phase position in the
plasma wakefield of the ionized electron (initially at rest), with
$\gamma_p$ is the Lorentz factor of the plasma wave phase velocity,
and $\phi$ is the potential of the wakefield (normalized to $m_e
c^2/e$), with $\phi_{\rm min}$ in the minimum amplitude of the
potential. Here we have assumed the wakefield near the axis is
approximately described by the 1D nonlinear wake potential
\cite{Esarey09}, whose extrema satisfy,
\begin{equation}
 \phi_{\rm min/max} = (E_z/E_0)^2/2 \pm \beta_p \left\{
\left[1+(E_z/E_0)^2/2\right]^2 -1 \right\}^{1/2}
,
\label{eq:phim}
\end{equation}
where $E_z/E_0$ is the peak of the accelerating wakefield normalized
to $E_0 = m_e c^2k_p/e$. The optimal ionization phase $\psi_i$ for
trapping is at the peak of the wake potential, $\phi (\psi_i) =
\phi_{\rm max}$ and $E_z(\psi_i)=0$. At this phase location, the
trapping condition is $1-\gamma_p^{-1} \leq \phi_{\rm max} - \phi_{\rm
min}$. In the limit $\gamma_p \gg 1$, the minimum wakefield amplitude
required for trapping is
\begin{equation}
 E_z/E_0 \geq (\sqrt{5}-2)^{1/2} \simeq 0.49 .
\end{equation} 
The longitudinal wakefield amplitude excited by a resonant,
circularly-polarized, Gaussian laser pulse is $E_z/E_0 \simeq
(\pi/2{\rm e})^{1/2} a_0^2 (1+a_0^2)^{-1/2}$, such that the required
laser amplitude for trapping is $a_0 > 0.94$.

Figure~\ref{fig:1d} illustrates an example of the concept of two-color
ionization injection. Figure~\ref{fig:1d}(a) shows the drive laser
vector potential $a_0$ (pulse centered at the origin), the field of
the excited plasma wake $E_z/E_0$ (red dashed curve), the ionization
laser vector potential [pulse centered at $k_p(z-ct)\simeq -9.5$] $a_i
\ll a_0$, and the trapping condition Eq.~\eqref{eq:trap-cond}, where
$0 \leq \phi - \phi_{\rm min} - 1 + 1/\gamma_p$ indicates an electron
ionized at rest will be trapped (blue curve). Figure~\ref{fig:1d}(b)
shows the electric field of the drive and ionization lasers
(normalized to the peak drive laser field), and the fraction of the
ionized gas state $f_i$ (red dashed curve). The normalized parameters
shown in Fig.~\ref{fig:1d} correspond to the case of a 10-$\mu$m
wavelength (CO$_2$) drive laser, with $a_0=2$ (linear polarization)
and 0.47~ps duration (FWHM laser intensity), propagating in a uniform
gas, producing a plasma with electron density $n_e =
10^{16}$~cm$^{-3}$, and resonantly exciting a wakefield. The
ionization pulse, with $a_i = 0.13$ (linear polarization), $\lambda_i
= 0.4$~$\mu$m (frequency-doubled Ti:Al$_2$O$_3$ laser), and 120~fs
duration (FWHM intensity), ionizes Kr$^{8+}$ to Kr$^{9+}$ (ionization
potential $U_i = 230$~eV) at a trapped wake phase.

\subsection{Trapped charge}

Tunneling ionization will determine the distribution of electrons
ionized, and, hence the trapped charge and emittance. The transverse
phase space distribution of the ionized electrons was derived in
Ref.~\citenum{Schroeder14}. The rms radius of the transverse
distribution of ionized electrons is \cite{Schroeder14}
\begin{equation}
\sigma_x = \sigma_y \simeq (w_i/\sqrt{2})\Delta ,
\label{eq:x}
\end{equation} 
where $w_i$ is the ionization laser spot size [i.e., $\bm{a}_i = a_i
\exp(-r^2/w_i^2) \hat{\bm{x}}$] and
\begin{equation}
\Delta 
 = \left( \frac{3\pi r_e a_i}{ \alpha^4 \lambda_i } \right)^{1/2} 
\left( \frac{U_H}{U_I} 
\right)^{3/4}
,
\label{eq:delta}
\end{equation} 
where $r_e = e^2/m_e c^2$ is the classical electron radius, $\alpha$
is the fine structure constant, and $U_i$ is the potential of the
state of the gas used for ionization injection, normalized to $U_H
\simeq 13.6$~eV. The parameter $\Delta^2$ is the normalized laser
field amplitude $E_i\propto a_i/\lambda_i$, and $\Delta^2 \ll 1$ is
satisfied at the ionization threshold. (Here we assume the Keldysh
parameter $\gamma_K = \left[2U_i/(m_ec^2a_i^2) \right]^{1/2}$
satisfies $\gamma_K <1$, such that tunneling ionization is the
dominant ionization mechanism.)

For an ionization laser pulse located at the proper phase $\psi_i$ of
a plasma wave of sufficient amplitude $E_z/E_0$, given by
Eq.~\eqref{eq:trap-cond}, electrons ionized at rest will be on trapped
orbits, and the amount of trapped charge will be determined by the
number of ionized electrons: $N_{t}= 2 \pi \sigma_{x}\sigma_{y}
\ell_{i} f_i n_{g}$, where $f_i$ is the ionization fraction, $n_g$ is
the number of ions with electrons in the proper charge state, and
$\ell_{i}$ is the interaction length. For interaction with a single
species gas, $n_g= n_e/Z$, where $n_e$ is the electron density and $Z$
is the number of electrons ionized by the drive laser (or
pre-ionized). If the length of the high-Z gas region is longer than
the ionization laser Rayleigh range, then the total charge will be
limited by the diffraction of the ionizing laser, $\ell_{i} \sim
Z_{Ri} = \pi w_{i}^{2}/\lambda_{i}$, and, hence, we expect the charge to
scale as $Q = eN_{t} \simeq e (\pi w_{i}^{2}\Delta^{2}) \ell_{i} f_i n_{g} 
\propto w_i^4$ with the interaction length limited by diffraction.
As shown below, the emittance scales as $\epsilon \propto w_i^2$, and,
hence, there is a trade-off between increasing the trapped charge
($N_t \propto w_i^4$, limited by diffraction) and reducing the
emittance.

\section{EMITTANCE} 

The achievable beam emittance using ionization injection into a
wakefield, driven by a charged particle beam or an intense laser, is
determined by the initial phase space distribution of the trapped
electrons and the plasma focusing forces.\cite{Schroeder14}. The rms
of the momentum distribution of the ionized electrons, in the plane of
laser polarization, is approximately \cite{Schroeder14}
\begin{equation}
 \sigma_{p_x} \simeq a_i \Delta ,
\label{eq:px}
\end{equation}
where $\Delta$ is given by Eq.~\eqref{eq:delta}. For a small
ionization injection laser amplitude $a_i^2\ll 1$, the momentum gain
from the ponderomotive force of the ionizing laser may be neglected.
Using Eqs.~\eqref{eq:x} and \eqref{eq:px}, the initial thermal
emittance of the beam generated by ionization injection is
\cite{Schroeder14}
\begin{equation}
 \epsilon_{t} = \left( \frac{3\pi r_e}{\sqrt{2} \alpha^4}\right) 
 \left( \frac{U_H}{U_i}\right)^{3/2} \frac{w_i a_i^2}{\lambda_i}
.
 \label{eq:e-therm}
 \end{equation}
Equation~\eqref{eq:e-therm} is independent of the driver of the plasma
accelerator (e.g., laser or particle beam) and is determined from the
ionization physics, i.e., the distribution of electrons produced via
tunneling ionization. Unless injected matched, the emittance of the
trapped beam will grow due to phase mixing in the plasma wakefield.

\subsection{Trapped particle dynamics}

An electron trapped in the plasma wakefield, i.e., ionized at the
proper phase to satisfy the trapping condition
Eq.~\eqref{eq:trap-cond}, will begin to accelerate and rotate in
transverse phase space. For an electron with initial transverse
position $x_0$ and normalized momentum $u_0$, the transverse position
and momentum will evolve as
\begin{align}
 x & \simeq \gamma^{-1/4}
\left[ x_0 \cos (k_\beta z) + (u_0/k_{\beta 0}) \sin (k_\beta
 z)\right] = \gamma^{-1/4} r_{\beta0} \cos (k_\beta z+ \varphi) 
,
\\
 u_x & = p_x/m_ec \simeq  \gamma^{1/4}
\left[ u_0 \cos (k_\beta z) - x_0 k_{\beta 0} \sin (k_\beta
 z)\right] = - \gamma^{1/4} k_{\beta 0} r_{\beta0} \sin (k_\beta z + 
\varphi) 
,
\end{align}
where $r_{\beta 0} = [x_0^2 + (u_0/k_{\beta 0})^2]^{1/2}$ is the
initial betatron amplitude, the energy $\gamma(z)$ is increasing as
the electron is accelerated in the wakefield,
$k_\beta\propto\gamma^{-1/2}$ is the betatron wavenumber in the
wakefield, and $k_{\beta 0} = k_\beta (z=0)$ is the initial betatron
wavenumber. For laser drivers with parameters satisfying $a_0
(1+a_0^2)^{-1/4} \gtrsim k_p w_0/2$ (assuming circular polarization),
or particle beam drivers, the driver will produce a co-moving ion
cavity with $k_\beta \simeq k_p /\sqrt{2\gamma}$.

Averaging the betatron amplitude over the initial beam distribution
yields \cite{Schroeder14}
\begin{equation}
\langle r_{\beta 0}^2 \rangle = \sigma_x^2 + \sigma_{px}^2/k_{\beta
0}^2 = \left[ 1 + \frac{2a_i^2}{(k_{\beta 0} w_i)^2} \right]
{w_i^2}\Delta^2/2
\label{eq:r_beta}
\end{equation}
in the plane of ionization laser polarization, and $\langle r_{\beta
0}^2 \rangle ={w_i^2}\Delta^2/2$ orthogonal to the plane of laser
polarization.

If the ionization injection region is short (due to a finite high-Z
gas region or short ionization injection pulse Rayleigh range),
followed by a post acceleration region that is sufficiently long to
allow betatron phase mixing, then the normalized emittance will
saturate to
\begin{equation}
\epsilon = \left[ \langle x^2 \rangle \langle u_x^2 \rangle -\langle x
u_x \rangle^2\right]^{1/2} = k_{\beta 0} \langle r_{\beta 0}^2 \rangle
/2 .
\label{eq:sat}
\end{equation}
The initial beam phase-space distribution Eq.~\eqref{eq:r_beta} will
determine the final saturated emittance Eq.~\eqref{eq:sat}. Matched
injection in the plane of laser polarization, where the thermal
emittance Eq.~\eqref{eq:e-therm} equals the saturated emittance
Eq.~\eqref{eq:sat}, occurs for laser-plasma parameters such that
$k_{\beta 0} w_i = \sqrt{2} a_i$. If continuous ionization injection occurs
over the length of the plasma accelerator, the saturated emittance
will asymptote to
$\epsilon =  k_{\beta 0} \langle r_{\beta 0}^2 \rangle /\sqrt{3}$.

\section{TWO-COLOR IONIZATION INJECTION EXAMPLE}
\label{sec:ex}

   \begin{figure}[t]
   \begin{center}
   \begin{tabular}{c}
 \begin{overpic}[]{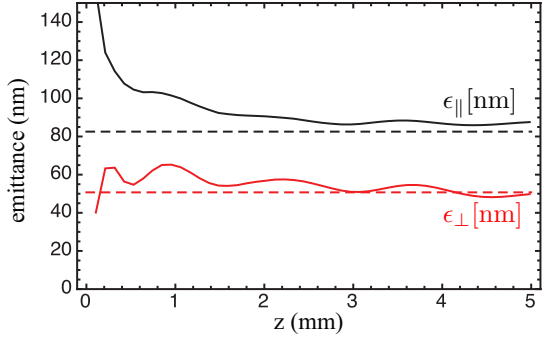}
 \put(81,43){$\epsilon_\parallel$[nm]}
 \put(81,22){\textcolor{red}{$\epsilon_\perp$[nm]}}
 \end{overpic}
   \end{tabular}
   \end{center}
   \caption[] 
{ \label{fig:CO2} Example of two-color ionization injection using
CO$_2$ drive laser: A (10 $\mu$m wavelength) 10~J, CO$_2$ laser with
$a=2$ (linear polarization), 0.47~ps duration (FWHM laser intensity),
and 155~$\mu$m spot size ($Z_R = 7.5$~mm) propagates in Krypton gas,
ionizing the gas up to Kr$^{8+}$, producing a plasma with electron
density 10$^{16}$~cm${^{-3}}$ and driving a wake. A frequency-doubled
(0.4~$\mu$m wavelength) Ti:Al$_2$O$_3$ laser with $a_i=0.14$ (linear
polarization), 120~fs duration (FWHM laser intensity), $w_i = 20~\mu$m
spot size ($Z_{Ri}=3$~mm), delayed with respect to the CO$_2$ laser,
ionizes Kr$^{8+}$ to Kr$^{9+}$ ($U_i=230$~eV) and generates a trapped
electron beam. Evolution of the transverse normalized emittance of the
trapped electron beam versus propagation distance $z$(mm) is plotted:
(black curve) emittance in the laser polarization plane from PIC
simulation, (dashed black line) $\epsilon_\parallel$ theoretical
saturated emittance in the laser polarization plane
Eq.~(\ref{eq:e-para}), (red curve) emittance orthogonal to the laser
polarization plane from PIC simulation, and (dashed red line)
$\epsilon_\perp$ theoretical saturated emittance orthogonal to the
laser polarization plane Eq.~(\ref{eq:e-perp}).}
   \end{figure} 

In this Section we present an example of beam generation using
two-color ionization injection with a short-pulse CO$_2$ laser driving
the wakefield and a frequency-doubled Ti:Al$_2$O$_3$ ionizing laser.
Modeling was done using the \textsc{inf\&rno} \cite{Benedetti10} and
\textsc{warp} \cite{Vay12} PIC codes of \textsc{blast} (Berkeley Lab
Accelerator Simulation Toolkit) \cite{blast}.
Figure~\ref{fig:CO2} shows the evolution of the transverse normalized
emittance of the trapped electron beam versus propagation distance
$z$. In Fig.~\ref{fig:CO2}, a 10 $\mu$m-wavelength, 10~J, CO$_2$ laser
with $a=2$ (linear polarization), 0.47~ps duration (FWHM laser
intensity), and 155~$\mu$m spot size ($Z_R = 7.5$~mm) propagates in
Krypton gas, ionizing the gas up to Kr$^{8+}$, producing a plasma with
electron density 10$^{16}$~cm${^{-3}}$ (gas density of $n_g
=1.25\times 10^{15}$~cm${^{-3}}$) and driving a wake. A
frequency-doubled (0.4~$\mu$m wavelength) Ti:Al$_2$O$_3$ laser with
$a_i=0.14$ (linear polarization), 120~fs duration (FWHM laser
intensity), $w_i = 20~\mu$m spot size ($Z_{Ri} = 3$~mm), delayed with
respect to the CO$_2$ laser, ionizes Kr$^{8+}$ to Kr$^{9+}$
($U_i=230$~eV) and generates a trapped electron beam. The solid black
and red solid curves are the normalized transverse emittance of the
trapped electron beam, parallel and orthogonal to the laser
polarization plane, respectively, calculated using the
\textsc{inf\&rno} module of \textsc{blast}.

For a small ionization injection laser amplitude $a_i^2\ll 1$,
propagating in a uniform gas jet, the final normalized transverse
emittances \cite{Schroeder14} are, using Eqs.~\eqref{eq:sat},
\eqref{eq:r_beta}, and \eqref{eq:delta},
\begin{align}
\epsilon_\parallel & = k_{\beta 0} w_i^2 
 \left[ 1 +   \frac{2a_i^2}{\left( k_{\beta 0} w_i\right)^2  }\right] 
\frac{a_i}{\lambda_i}
\left( \frac{3 \pi r_e}{4 \alpha^4}\right) 
\left( \frac{U_H}{U_i}\right)^{3/2} ,
\label{eq:e-para}
\\
\epsilon_\perp & = k_{\beta 0} w_i^2 
\frac{a_i}{\lambda_i}
\left( \frac{3 \pi r_e}{4 \alpha^4}\right) 
\left( \frac{U_H}{U_i}\right)^{3/2} ,
\label{eq:e-perp}
\end{align}
where $\epsilon_\parallel$ ($\epsilon_\perp$) is the transverse
normalized emittance parallel (orthogonal) to the plane of ionization
laser polarization. Shown in Fig.~\ref{fig:CO2} are the theoretical
predictions: (dashed black line) $\epsilon_\parallel$ emittance in the
laser polarization plane Eq.~(\ref{eq:e-para}) and (dashed red line)
$\epsilon_\perp$ emittance orthogonal to the laser polarization plane
Eq.~(\ref{eq:e-perp}). For these drive laser-plasma parameters,
$k_{\beta 0}/(k_p/\sqrt{2}) \simeq 0.9$. Good agreement is achieved
between the PIC simulation and the theoretical estimates
Eqs.~(\ref{eq:e-para}) and (\ref{eq:e-perp}). Phase mixing is not
complete after 5~mm, and the emittance orthogonal to the plane of
laser polarization is still varying.
The charge contained in the beam modeled with the PIC code is 27~pC.
This approximately agrees with the order-of-magnitude estimate of the
trapped charge over a Rayleigh length, assuming $f_i \approx 1$,
\begin{equation}
 Q = eN_t \sim e n_g
Z_{Ri} \pi w_i^2 \Delta^2 
,
\label{eq:charge}
\end{equation}
which, for the parameters of Fig.~\ref{fig:CO2}, yields $Q\simeq
34$~pC.

\begin{figure}[]
    \begin{center}
    \begin{tabular}{c}
\begin{overpic}[]{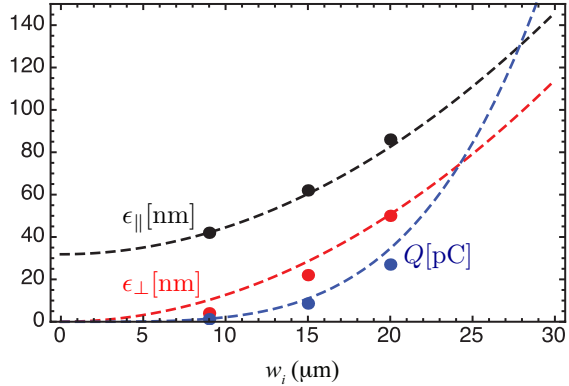}
 \put(20,30){$\epsilon_\parallel$[nm]}
 \put(20,18){\textcolor{red}{$\epsilon_\perp$[nm]}}
\put(70,23){\textcolor{DarkBlue}{$Q$[pC]}}
 \end{overpic}
    \end{tabular}
    \end{center}
    \caption[] 
{ \label{fig:rs} Emittance [parallel, Eq.~\eqref{eq:e-para} (dashed
black curve), and orthogonal, Eq.~\eqref{eq:e-perp} (dashed red
curve), to the ionization laser polarization] and the charge
Eq.~\eqref{eq:charge} (dashed blue curve) versus ionization laser spot
size (all other parameters are the same as Fig.~\ref{fig:CO2}). Points
are PIC simulation results for the emittance parallel
$\epsilon_\parallel$[\textrm{nm}] (black) and orthogonal
$\epsilon_\perp$[\textrm{nm}] (red) to the laser polarization, and the
charge $Q[\textrm{pC}]$ (blue).}
    \end{figure}

In general, a trade-off exists between the reducing the emittance
($\epsilon \propto w_i^2$) and increasing the trapped charge ($Q
\propto w_i^4$, with interaction length limited by diffraction).
Figure~\ref{fig:rs} shows the theoretical estimates for the beam
emittance [parallel, Eq.~\eqref{eq:e-para} (dashed black curve), and
orthogonal, Eq.~\eqref{eq:e-perp} (dashed red curve), to the
ionization laser polarization] and the charge, Eq.~\eqref{eq:charge}
(dashed blue curve) versus ionization laser spot size (all other
parameters are the same as Fig.~\ref{fig:CO2}). The points in
Fig.~\ref{fig:rs} are PIC simulation results (calculated using the
\textsc{inf\&rno} module of \textsc{blast}) for the emittance in the
laser polarization plane $\epsilon_\parallel [\textrm{nm}]$ (black)
and orthogonal to the laser polarization plane
$\epsilon_\perp$[\textrm{nm}] (red), and the charge $Q[\textrm{pC}]$
(blue).

%

\section{SUMMARY AND CONCLUSIONS}

Ultra-low emittance electron beams, on the order of tens of nm, can be
generated in a plasma accelerator using ionization injection in a
wakefield \cite{Hidding12,Yu14,Schroeder14}. In this paper we have
reviewed the trapping physics, distribution of ionized electrons, and
the trapped beam dynamics. Expressions for the achievable beam
emittance using ionization injection into a plasma wakefield were
presented.\cite{Schroeder14} Note that these results are independent
of the wakefield driver (laser \cite{Yu14} or particle beam
\cite{Hidding12}). An experimentally-relevant example was presented of
ultra-low emittance beam generation by two-color ionization injection
using a short-pulse, CO$_2$ laser driving the wakefield and a
frequency-doubled, Ti:Al$_2$O$_3$ laser ionizing a Krypton gas. In
this example the interaction region was limited by the Rayleigh range
of the ionizing laser. In general, a trade-off exists between the
reducing the saturated emittance ($\epsilon \propto w_i^2$) and
increasing the trapped charge ($N_t \propto w_i^4$, with interaction
length limited by diffraction). Self-consistent numerical modeling of
the beam injection for this example was performed and compared to
analytic estimates, showing good agreement. It should be noted that
this example is not fully optimized, and, depending on the
laser-plasma parameters, other gases may be considered (e.g., Ne) for
improved performance.


\acknowledgments    
 
This work was supported by the Director, Office of Science, Office of
High Energy Physics, of the U.S. Department of Energy under Contract
Nos.~DE-AC02-05CH11231. This research used computational resources of
the National Energy Research Scientific Computing Center, which is
supported by the Office of Science of the U.S. Department of Energy
under Contract No.~DE-AC02-05CH11231. This work was supported in part
by the National Basic Research Program of China (Grant
No.~2013CBA01504) and the National Natural Science Foundation of China
(Grant No.~11405107).



\begin{thebibliography}{10}

\bibitem{Esarey09}
E.~Esarey, C.~B. Schroeder, and W.~P. Leemans, ``Physics of laser-driven
  plasma-based electron accelerators,'' {\em Rev. Mod. Phys.}~{\bf 81},
  pp.~1229--1285, July--September 2009.

\bibitem{Wang13}
X.~Wang, R.~Zgadzaj, N.~Fazel, Z.~Li, S.~A. Yi, X.~Zhang, W.~Henderson, Y.-Y.
  Chang, R.~Korzekwa, H.-E. Tsai, C.-H. Pai, H.~Quevedo, G.~Dyer, E.~Gaul,
  M.~Martinez, A.~C. Bernstein, T.~Borger, M.~Spinks, M.~Donovan, V.~Khudik,
  G.~Shvets, T.~Ditmire, and M.~C. Downer, ``Quasi-monoenergetic laser-plasma
  acceleration of electrons to {2 GeV},'' {\em Nature Communications}~{\bf 4},
  p.~1988, June 2013.

\bibitem{Leemans14}
W.~P. Leemans, A.~J. Gonsalves, H.-S. Mao, K.~Nakamura, C.~Benedetti, C.~B.
  Schroeder, C.~T\'oth, J.~Daniels, D.~E. Mittelberger, S.~S. Bulanov, J.-L.
  Vay, C.~G.~R. Geddes, and E.~Esarey, ``Multi-{GeV} electron beams from
  capillary-discharge-guided subpetawatt laser pulses in the self-trapping
  regime,'' {\em Phys. Rev. Lett.}~{\bf 113}, p.~245002, December 2014.

\bibitem{Benedetti13}
C.~Benedetti, C.~B. Schroeder, E.~Esarey, F.~Rossi, and W.~P. Leemans,
  ``Numerical investigation of electron self-injection in the nonlinear bubble
  regime,'' {\em Phys. Plasmas}~{\bf 20}, p.~103108, October 2013.

\bibitem{Plateau12}
G.~R. Plateau, C.~G.~R. Geddes, D.~B. Thorn, M.~Chen, C.~Benedetti, E.~Esarey,
  A.~J. Gonsalves, N.~H. Matlis, K.~Nakamura, C.~B. Schroeder, S.~Shiraishi,
  T.~Sokollik, J.~{van Tilborg}, {Cs. T\'{o}th}, S.~Trotsenko, T.~S. Kim,
  M.~Battaglia, T.~St\"{o}hlker, and W.~P. Leemans, ``Ultra-low-emittance
  electron bunches from a laser-plasma accelerator measured using single-shot
  x-ray spectroscopy,'' {\em Phys. Rev. Lett.}~{\bf 109}, p.~064802, August
  2012.

\bibitem{Weingartner12}
R.~Weingartner, S.~Raith, A.~Popp, S.~Chou, J.~Wenz, K.~Khrennikov,
  M.~Heigoldt, A.~R. Maier, N.~Kajumba, M.~Fuchs, B.~Zeitler, F.~Krausz,
  S.~Karsch, and F.~Gr\"uner, ``Ultralow emittance electron beams from a
  laser-wakefield accelerator,'' {\em Phys. Rev. ST Accel. Beams}~{\bf 15},
  p.~111302, November 2012.

\bibitem{Hidding12}
B.~Hidding, G.~Pretzler, J.~B. Rosenzweig, T.~K\"onigstein, D.~Schiller, and
  D.~L. Bruhwiler, ``Ultracold electron bunch generation via plasma
  photocathode emission and acceleration in a beam-driven plasma blowout,''
  {\em Phys. Rev. Lett.}~{\bf 108}, p.~035001, January 2012.

\bibitem{Yu14}
L.-L. Yu, E.~Esarey, C.~B. Schroeder, J.-L. Vay, C.~Benedetti, C.~G.~R. Geddes,
  M.~Chen, and W.~P. Leemans, ``Two-color laser-ionization injection,'' {\em
  Phys. Rev. Lett.}~{\bf 112}, p.~125001, March 2014.

\bibitem{Yu13}
L.-L. Yu, E.~Esarey, J.-L. Vay, C.~B. Schroeder, C.~Benedetti, C.~G.~R. Geddes,
  S.~G. Rykovanov, S.~S. Bulanov, M.~Chen, and W.~P. Leemans, ``Low transverse
  emittance electron bunches from two-color laser-ionization injection,'' in
  {\em Proc. SPIE},   {\bf 8779}, p.~877908, SPIE Digital Library, May 2013.

\bibitem{Xu14b}
X.~L. Xu, Y.~P. Wu, C.~J. Zhang, F.~Li, Y.~Wan, J.~F. Hua, C.-H. Pai, W.~Lu,
  P.~Yu, C.~Joshi, and W.~B. Mori, ``Low emittance electron beam generation
  from a laser wakefield accelerator using two laser pulses with different
  wavelengths,'' {\em Phys. Rev. ST Accel. Beams}~{\bf 17}, p.~061301, June
  2014.

\bibitem{Schroeder14}
C.~B. Schroeder, J.-L. Vay, E.~Esarey, S.~S. Bulanov, C.~Benedetti, L.-L. Yu,
  M.~Chen, C.~G.~R. Geddes, and W.~P. Leemans, ``Thermal emittance from
  ionization-induced trapping in plasma accelerators,'' {\em Phys. Rev. ST
  Accel. Beams}~{\bf 17}, p.~101301, October 2014.

\bibitem{Pogorelsky14}
I.~V. Pogorelsky and I.~{Ben-Zvi}, ``{Brookhaven National Laboratory's
  Accelerator Test Facility: Research Highlights and Plans},'' {\em Plasma
  Phys. Control. Fusion}~{\bf 56}, p.~084017, July 2014.

\bibitem{MinChen12}
M.~Chen, E.~Esarey, C.~B. Schroeder, C.~G.~R. Geddes, and W.~P. Leemans,
  ``Theory of ionization-induced trapping in laser-plasma accelerators,'' {\em
  Phys. Plasmas}~{\bf 19}, p.~033101, March 2012.

\bibitem{Benedetti10}
C.~Benedetti, C.~B. Schroeder, E.~Esarey, C.~G.~R. Geddes, and W.~P. Leemans,
  ``Efficient modeling of laser-plasma accelerators with {INF\&RNO},'' in {\em
  Proc. of 2010 Advanced Accelerator Concepts Workshop},  G.~Nusinovich and
  S.~Gold, eds.,  {\bf 1299}, pp.~250--255, AIP, (NY), 2010.

\bibitem{Vay12}
J.-L. Vay, D.~P. Grote, R.~H. Cohen, and A.~Friedman, ``Novel methods in the
  particle-in-cell accelerator code-framework warp,'' {\em Comput. Sci.
  Discovery}~{\bf 5}, p.~014019, 2012.

\bibitem{blast}
``{Berkeley Lab Accelerator Simulation Toolkit}.'' {http://blast.lbl.gov}.

\end{thebibliography}

\end{document}